----------------------------------------------------------------------- 
\documentclass[eqsecnum,aps,twocolumn]{revtex4} 

\begin{document}


%
\title{Low-energy muon-transfer reaction  from
hydrogen isotopes 
to helium  isotopes}

\author{Renat
A. Sultanov$^{a,b,}$\thanks{e-mail: rsultanov@pacificcoast.net} and Sadhan
K. Adhikari$^{a,}$\thanks{e-mail: adhikari@ift.unesp.br}}

\address{$^a$Instituto de F\'\i sica Te\'orica,
Universidade Estadual Paulista,
01405-900 S\~{a}o Paulo, S\~{a}o Paulo, Brazil\\
$^b$ \#1-10295-128A Street,
           Surrey, B.C. V3T 3E7
           Canada}

\date{\today}

\begin{abstract}

Direct muon transfer in low-energy collisions of the muonic hydrogen
H$_\mu$ and helium (He$^{++}$) is considered in a three-body
quantum-mechanical framework of coordinate-space integro-differential
Faddeev-Hahn-type equations within two- and six-state close coupling
approximations.  The final-state Coulomb interaction is treated without
any approximation employing appropriate Coulomb waves in the final state.
The present results agree reasonably well with previous semiclassical
calculations.
 
{PACS number(s): 36.10.Dr}

\end{abstract}

\maketitle

\vskip1.5pc
%
%

Experimental investigation of the low-energy muon
transfer in
collision of muonic hydrogen H$_\mu$ (bound state of a
hydrogen isotope
and muon $\mu^-$) and He is important for muon
catalyzed fusion cycle
of hydrogen isotopes \cite{Rafelski-1991}. If the
hydrogen is 
contaminated
by He, then the muon-catalyzed fusion of hydrogen
isotopes will be
affected by muon transfer from H$_\mu$ to He. This
makes the study of 
muon
transfer from H$_\mu$ to He of special relevance. The
study of such
collisions involving three charged particles is also 
interesting from
a theoretical point of view as an example of
rearrangement scattering 
with
Coulomb interaction in the final state.  Such
reactions with
post-collision Coulomb interaction between clusters
appear frequently 
in
atomic and molecular physics.

Recently, we formulated \cite{ska1} a few-body
quantum-mechanical
description of
direct muon-transfer reaction from H$_\mu$ to
different nuclei using 
the
Faddeev-Hahn-type few-body equations
\cite{Faddeev-1961,Hahn-1968} in 
the
close-coupling approximation.  
We applied this formulation successfully to the study
of muon transfer
among various hydrogen isotopes. In that study there
was no final-state
Coulomb interaction.  Later on we extended this
approach to the study 
of
muon transfer from hydrogen isotopes to He and Li,
where the 
final-state
Coulomb interaction is appropriately included in the
calculational 
scheme 
\cite{ska4}.  
Next we included heavier targets such as, C, O
\cite{ska4,ska3}, Ne, S,
and Ar
\cite{ska5}, in the scheme, where appropriate
polarization potentials 
were
included in addition to the final-state Coulomb
interaction. In case of
these heavy targets the muon transfer rates from
H$_\mu$ agree well 
with
the existing experimental results
\cite{ska4,ska3,ska5}. In case of 
these
heavier
targets there
is evidence \cite{ska5,xyz} that the muon is absorbed
in a specific 
final
state of the
heavy nuclei, and hence, the muon-transfer rates were
calculated in a
two-state close-coupling approximation including a
single state each of
the target and the incident H$_\mu$.

In this report, we perform a two- and six-state
calculation of direct
muon-transfer from H$_\mu$ to He isotopes and compare
with other
calculations. The reaction we study is 
\begin{equation}
({\mbox H}_\mu)_{1s}+{\mbox He}^{++} \to ({\mbox
He}_\mu)_{1s}^+
+{\mbox
H}^+,
\end{equation}
where ${\mbox He}^{++}$ represents the He nucleus and
$ ({\mbox
He}_\mu)_{1s}^+ $ the hydrogen-like bound state of 
${\mbox He}^{++}$ 
and
a muon.  The present study is meant to test the
accuracy
and the convergence trend of our approach.
For the lighter target such as He, muon transfer could
take
place via direct as well as compound molecular
formation. The 
experimental
rates include the molecular formation
\cite{Tresch1-1998}.  The method 
we
consider here is not applicable to muon transfer via
molecular 
formation.
The experimental transfer rates via molecular
formation leads to much
higher results in the case of target He while compared
to the 
theoretical
result of direct muon transfer. For heavier targets,
such as C, O, Ne, 
Ar
etc., the molecular rates are highly suppressed and 
the muon transfer takes place essentially via the
direct channel and
our theoretical direct rates are in agreement with
experiment
\cite{ska4,ska3,ska5}.

For the theoretical treatment of a  three-body
muon-transfer  
rearrangement process,
Faddeev-type equations \cite{Faddeev-1961}, especially
the modified
version proposed by Hahn \cite{Hahn-1968}, appear to
be
very suitable. 
The  two possible 
asymptotic two-cluster 
configurations of the above rearrangement problem  are
conveniently tackled by a set
of two coupled Faddeev-Hahn-type
equations for components $\Psi_1$ and $\Psi_2$ of the
wave function 
$\Psi
= \Psi_1 + \Psi_2$, where each component carrys the
asymptotic boundary condition for a specific
configuration
\cite{renat}.
These equations are very useful to
incorporate distortion potentials for specific initial
and final 
asymptotic states \cite{Hahn-1972}.
It is
possible to include the final-state Coulomb
interaction explicitly in
these
equations, so that a low-order approximation to these
equations 
produces 
the correct asymptotic behavior \cite{Hahn-1972}.

 Here we quote  the  Faddeev-Hahn-type two-component
equations used in 
the
close coupling type approximation 
for the calculation of muon transfer rates
\cite{ska1,ska3}.
We denote 
He$^{++}$ by ${\sf 1}$, the hydrogen isotope(s)
by ${\sf 2}$ and muon by ${\sf 3}$.
Below the three-body breakup threshold, following
two-cluster asymptotic configurations
are possible in the system {\sf 123}:  $({\sf 23})\ -\
{\sf 1}$ and
$({\sf 13})\ -\ {\sf 2}$. These two configurations
  correspond to two distinct physical channels, 
denoted by
1 and 2, respectively.
These configurations
are  determined by the Jacobi coordinates
$(\vec r_{j3}, \vec \rho_k)$: $\vec r_{13} = \vec r_3
- \vec r_1,
\hspace{6mm} \vec \rho_2 =
(\vec r_3 + m_1\vec r_1) / (1 + m_1) - \vec r_2$,
$\vec r_{23} = \vec r_3 - \vec r_2,
\hspace{6mm} \vec \rho_1 =
(\vec r_3 + m_2\vec r_2) / (1 + m_2) - \vec r_1$,
where
$\vec r_{i}$, $m_{i}$ ($i=1, 2, 3,$) are coordinates
and
masses of the particle $i$, respectively.

Let us introduce
the total three-body wave function as a sum of two
components
\cite{Hahn-1968}
\begin{equation}
\Psi(\vec r_1, \vec r_2, \vec r_3) \ =\  \Psi_1 (\vec
r_{23},\vec 
\rho_1)
\ + \ \Psi_2 (\vec r_{13},\vec \rho_2)
\label{eq:total2}
\end{equation}
where $\Psi_1 (\vec r_{23},\vec \rho_1)$
is quadratically integrable over the variable
$\vec r_{23}$, and  $\Psi_2 (\vec r_{13},\vec \rho_2)$
over
$\vec r_{13}$. The components $\Psi_1$ and $\Psi_2$
carry the
asymptotic
boundary condition for channels 1 and 2, respectively.
The second component is responsible for
pure Coulomb interaction in the final state.
These components satisfy the following
set of two coupled equations
\cite{ska3}\begin{eqnarray}\label{eq:1a}
[E &-& (H_0 + V_{23}(\vec r_{23})) ]\Psi_1 (\vec
r_{23}, \vec \rho_1)  \nonumber \\
&=&[(V_{23}({\vec r_{23}})  + V_{12}({\vec r_{12}})) 
 -
U_C (\vec \rho_2)]
\Psi_2 (\vec r_{13}, \vec \rho_2)\;
\end{eqnarray}
\begin{eqnarray}
[E &-& (H_0 + V_{13}(\vec r_{13})) - \ U_{  C}\ (\vec
\rho_2)]\Psi_2 (\vec r_{13}, \vec \rho_2)   \nonumber
\\       
&=& [(V_{13} (\vec r_{13})+ V_{12}(\vec r_{12}))
]\Psi_1 (\vec r_{23}, \vec \rho_1)
\label{eq:1} 
\end{eqnarray}
where $E$ is the center-of-mass energy, $H_0$ the
total kinetic energy
operator,  $V_{ij} (\vec r_{ij})$
the pair potential $(i \not= j = 1, 2, 3)$,
$U_{ C}$ the final-state Coulomb  interaction.
For both $U_C=0$ and $U_C\ne 0$, Eqs. \ref{eq:1a}
and \ref{eq:1} are equivalent to the underlying
Schr\"odinger equation for the problem. Hence a
converged solution of these equations should lead to
the true dynamical solution of the problem. However,
an approximate (low-order) close-coupling solution of
these equations with the proper final-state Coulomb
interaction $U_C$ should lead to a more converged
solution when compared with a similar solution with
$U_C=0$ as we shall see in the following.  At higher energies the effect
of the explicit inclusion of the Coulomb potential  $U_c$ in the equations
above is reduced  and the transfer rates for $U_c=0$ tends towards those
for $U_c\ne 0$.

We solve the integro-differential form of the
Faddeev-Hahn equation by
the close-coupling approximation scheme involving up
to six states. 
This
procedure consists in expanding the wave function
components $\Psi_1$ 
and
$\Psi_2$ in terms of eigenfunctions of subsystem
Hamiltonians in 
initial
and final channels, respectively.  Although, these
subsystem
eigenfunctions are not orthogonal to each other, the
components 
$\Psi_1$
and $\Psi_2$ satisfy a coupled set of equations
incorporating the 
correct
asymptotic behavior of the wave function.
Consequently, there is no
problem of overcompleteness as encountered in similar
expansion 
approaches
for rearrangement reactions based on the Schr\"odinger
equation. The
resultant coupled Faddeev-Hahn-type equations are then
projected on the
expansion functions. After a partial-wave projection
this leads to a 
set
of one-dimensional coupled integro-differential
equations for the
expansion coefficients, which is solved numerically.
The
mathematical details of the
approach have appeared elsewhere and we refer the
interested readers to
the original references \cite{ska1,ska4}.

First, we restrict ourselves to a two-level
approximation by choosing
in the relevant close-coupling expansion  the
hydrogen-like
ground states $(\mbox{H}_\mu)_{1s}$ and
$({\mbox{He}}_\mu)^{+}_{1s}$,
where H $={^1\mbox{H}} ^+$ and $^2\mbox{H}^+$, 
and He =
$^3{\mbox{He}}^{++}$
and
$^4{\mbox {He}}^{++}$. 
Numerically, stable and converged results were obtained
in these cases. 
The rates $\lambda_{\mbox{tr}}$ $/10^6$ sec$^{-1}$
for both $U_C=0$ and $U_C\ne 0$
at low energies are presented in Table I
together with the results of Refs. 
\cite{W-1992,Matv-1973,Sultanov-1999}.
Next we extend the calculation to  the
six-state
model where we include the  H$_\mu$(1s,2s,2p)  and 
(He$_\mu)^+$(1s,2s,2p) 
states of both muonic hydrogen and
helium. These transition rates  for both $U_C=0$ and
$U_C\ne 0$ 
are also shown in Table I. All rates 
have converged to the precision shown in this 
table.
The present results are consistent with the
phenomenological 
isotope effect, e.g., the rate
decreases from $^1$H to $^2$H \cite{iso2}. 
The effect of including the (2s,2p) states in the
calculational
scheme is also explicit.

The present six-state calculation, when compared with the two-state
calculation, shows the trend of convergence. We see
that at high energies the $U_c\ne 0$ rates tend towards the $U_c=0$
rates. This is illustrated in Table I for the two-state model for two
systems.
However, the effect of the inclusion of $U_c$ is significant at low
energies. The two-state rates with $U_c=0$ are the least converged. The six-state rates with $U_c=0$ show the direction of
convergence.  From an analysis of our results reported in Table I, we find, as
expected, the six-state results with $U_c\ne 0$ are the most converged.
The two-state rates  with $U_c\ne 0$ are more converged than the
three-state rates with $U_c=0$. 
The explicit inclusion of the
final-state Coulomb interaction $U_C$ has enhanced the
convergence and has led to more consistent rates.  
The
present six-state results for $U_C\ne 0$ are very
close to the semiclassical results 
of
Ref.
\cite{W-1992} for all four systems considered.
However, the agreement 
with the results of 
Ref.  \cite{Matv-1973}
is poor; the present rates for $^1$H$_\mu$
($^2$H$_\mu$) are roughly
double
(ten times) the rates of Ref. \cite{Matv-1973}.
Considering the very
qualitative and semiclassical nature of the
calculation of
Ref. \cite{Sultanov-1999}, the fair agreement with the
present few-body
quantum  
treatment and with Ref. \cite{W-1992} is encouraging.

The study of three-body charge transfer reactions with Coulomb repulsion
in the final state has been the subject of this report. We have studied
such reactions employing a detailed few-body description of the
rearrangement scattering problem by solving the Faddeev-Hahn-type
equations \cite{Faddeev-1961,Hahn-1968} in coordinate space. To provide
correct asymptotic form in the final state the pure Coulomb interaction
has been incorporated directly into the equations. It is shown that within
this formalism, the application of a close-coupling-type ansatz leads to
satisfactory results already in low-order approximations for direct
muon-transfer reactions between hydrogen isotopes and light nuclei ${\mbox
{He}}^{++}$.  Because of computational difficulties, in this application
we have considered up to six states in the expansion scheme (1s,2s,2p on
each center $-$ H$_\mu$ and He$_\mu^+$), which may not always be adequate.
Further calculations with larger basis sets are needed to obtain accurate
converged results. However, the inclusion of three basis states on each
center is expected to build in a satisfactory account of the polarization
potential in the model and should lead to physically acceptable results.
The agreement of the present rates with those of the semiclassical model
also assuring.


\acknowledgments
We acknowledge the financial support from
Funda\c{c}\~{a}o
de Amparo \~{a} Pesquisa do Estado de S\~{a}o Paulo of
 Brazil.
The numerical calculations have been performed on the
IBM SP2
Supercomputer of the Departamento de
F\'\i sica - IBILCE - UNESP,
S\~{a}o Jos\'e do Rio Preto, Brazil.


\mediumtext 
\begin{table}
{Table I.
Low-energy muon-transfer rates
$\lambda_{\mbox{tr}}$$/10^6$
$\mbox{sec}^{-1}$
from proton  (${^1{\mbox H}_\mu}$)$_{1s}$
and deuteron (${^2{\mbox H}_\mu}$)$_{1s}$
to hydrogen-like ground states
(${^3{\mbox {He}}}_\mu$)$^+_{1s}$ and  
(${^4{\mbox {He}}}_\mu$)$^+_{1s}$
within six-state close-coupling model.}
\begin{tabular}{lcccccccccccccccc}
\hline
\multicolumn{1}{l}{System}                &
\multicolumn{1}{c}{Energy}                     &
\multicolumn{2}{c}{Two-state result} &
\multicolumn{2}{c}{Six-state result} &
\multicolumn{1}{c}{Other result}\\
\multicolumn{1}{l}{}              &
\multicolumn{1}{l}{$(eV)$}                &
\multicolumn{1}{c}{$U_c=0$} &
\multicolumn{1}{c}{$ U_c\ne 0$} &
\multicolumn{1}{c}{$U_c=0$} &
\multicolumn{1}{c}{$U_c\ne 0$} &
\multicolumn{1}{c}{}  &
\\ \hline
\hline
$ {^1{\mbox H}_\mu} + {^3{\mbox {He}}^{++}} $
& $ 0.001 $ & 2.1 & 8.4 & 3.8&13.6 
 & $ $ & $ $\\
$ $
& $ 0.01 $ & 2.1 & 8.4 & 3.4& 13.3
  \\
$ $
& $ 0.04 $ & 2.1 & 8.4 & 3.3&
12.5&10.9\cite{W-1992},6.3\cite{Matv-1973},7.25\cite{Sultanov-1999}   
\\
$ $
& $ 0.1 $ & 2.0 & 8.3 & 3.2 & 11.0
\\
$ $
& $ 5 $ & 1.4 & 5.6 & 
& \\
$ $
& $ 60 $ & 0.5 & 1.0 & 
& \\
\hline
$ {^2{\mbox H}_\mu} + {^3{\mbox {He}}^{++}} $
& $ 0.001 $ & 1.3 & $5.2 $ & 2.8&9.7
& $ $ & $ $ & $ $\\
$ $
& $ 0.01 $ & 1.3 & $ 5.2 $ &2.7 &9.5
\\
$ $
& $ 0.04 $ & 1.3& $ 5.2 $ &2.6 &9.4&
9.6\cite{W-1992},
1.3\cite{Matv-1973},
4.77\cite{Sultanov-1999}\\
$ $
& $ 0.1 $ & 1.2 & $ 5.1 $ &2.3 &8.8
\\
$ $
& $ 5 $ &0.6 & 3.2 & 
\\
$ $
& $ 60 $ & 0.1 & $0.15  $ & 
\\
\hline
$ {^1{\mbox H}_\mu} + {^4{\mbox {He}}^{++}} $
& $ 0.001 $ &1.4  & 6.8 &3.2 &12.8
\\
$ $
& $ 0.01 $ & 1.4& $ 6.8 $ &2.8 &12.4
\\
$ $
& $ 0.04 $ &  1.4& $ 6.8 $ &2.6 &11.8&
10.7\cite{W-1992},5.5\cite{Matv-1973},6.65\cite{Sultanov-1999}\\
$ $
& $ 0.1 $ &  1.3& $ 6.7 $ & 2.3&10.6
\\
\hline
$ {^2{\mbox H}_\mu} + {^4{\mbox {He}}^{++}} $
& $ 0.001 $ & 0.8 & $5.0 $ &1.7 &9.6
& $ $ & $ $ & $ $\\
$ $
& $ 0.01 $ & 0.8 & $ 5.0 $  &1.6 &9.5
&\\
$ $
& $ 0.04 $ & 0.8 & $ 5.0 $ &1.6 &9.3
&9.6\cite{W-1992},1.0\cite{Matv-1973},4.17\cite{Sultanov-1999}     \\
$ $
& $ 0.1 $ & 0.8 & $ 4.9 $ & 1.2&8.8
& \\
\hline
\end{tabular}
\end{table}

\end{document}